\documentclass[11pt]{article}
\usepackage{amssymb}
\usepackage{amsmath}
\usepackage[dvips]{graphicx} 
\begin{document}

\begin{center}

{\Large On metallic gratings coated conformally with isotropic
negative--phase--velocity materials}\\
\vskip 1cm

{ \bf Marina E. Inchaussandague\footnote{Corresponding Author. E-mail: mei@df.uba.ar}$^{(a,b)}$, Akhlesh  Lakhtakia\footnote{E-mail: akhlesh@psu.edu}$^{(c)}$ and
Ricardo A. Depine\footnote{E-mail: rdep@df.uba.ar}$^{(a,b)}$ 
} \\
\bigskip

{$^{(a)}$ GEA~---~Grupo de Electromagnetismo Aplicado, Departamento de F\'{\i}sica,}\\
{Facultad de Ciencias Exactas y Naturales, Universidad de Buenos Aires,}\\
{Ciudad Universitaria, Pabell\'{o}n I, 1428 Buenos Aires, Argentina.}\\
{$^{(b)}$ CONICET~---~Consejo Nacional de Investigaciones Cient\'{\i}ficas y T\'ecnicas,}\\
{Rivadavia 1917, Buenos Aires, Argentina.}\\
{$^{(c)}$ CATMAS~---~Computational and Theoretical Materials Sciences Group,\\
Department of Engineering Science and
Mechanics,\\ Pennsylvania State University, University Park, PA
16802--6812, USA.}\\

\end{center}

\bigskip
\noindent ABSTRACT--Application of the differential method (also called the C method) 
to plane--wave diffraction by a perfectly conducting, sinusoidally corrugated metallic grating coated with a linear, homogeneous, isotropic, lossless dielectric--magnetic  material shows that coating materials with negative index of refraction 
may deliver enhanced maximum nonspecular reflection efficiencies in comparison
to coating materials with positive index of refraction.

\bigskip


\bigskip

\section{Introduction}
Surface--relief gratings are commonly made of metals. Often, dielectric coatings are deposited on top
of such gratings for protection against environmental degradation as well as
to increase the reflection efficiency (which is the same as
the specular reflectance) \cite{SS85,PMM}. Thus, a coated grating
is a dielectric slab sitting atop a periodically corrugated metallic 
surface. Various different types of dielectric coatings have been examined,
ranging from conformal coatings to coatings that simply fill up the
grooves of the gratings \cite{CA70,HVM74,LVV91}.

The recent emergence of electromagnetic metamaterials \cite{Walser} has opened
new avenues. Metamaterial coatings could be employed
for tailoring the specular and the nonspecular reflectances of
metallic gratings.  Specifically, metamaterials wherein the time--averaged
Poynting vector is in opposition to the phase velocity are very attractive \cite{LMW,SAR}.
Indeed, gratings made of isotropic negative--phase--velocity (NPV) materials
have been analytically investigated already \cite{DLoc,DLpre,DLS}, and numerical results
show that they function very differently from their positive--phase--velocity (PPV)
analogs.

The focus of this communication is on sinusoidally corrugated metallic gratings
with conformal coatings of constant thickness made of 
homogeneous, isotropic NPV materials. The differential or the C method
\cite{chande,madeeasy,chandea2,DLopt} was adapted for
that purpose. A comparison of NPV and PPV coating materials
in relation to specular and nonspecular reflection
efficiencies was undertaken.

\section{Diffraction problem}
\begin{center} 
\begin{figure}[ht] 
\begin{tabular}{c}
\includegraphics[width=20cm]{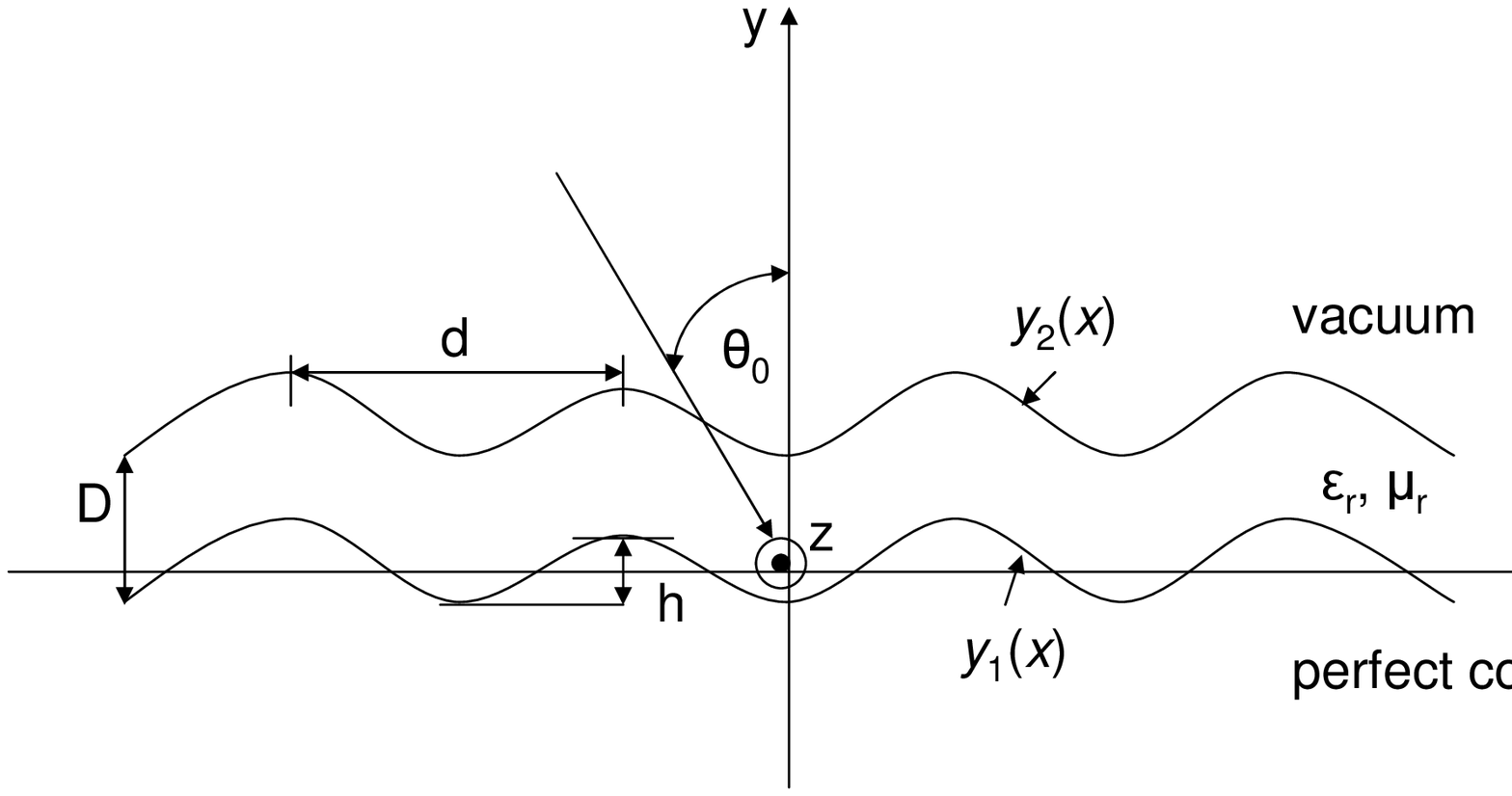} 
\end{tabular}
\caption{Geometry of the diffraction problem.}
\label{Fig1}
\end{figure}
\end{center}

Let us consider plane wave diffraction from a coated perfectly conducting  grating illuminated by a plane wave. The geometry of the diffraction problem is shown in Figure~\ref{Fig1}. The region of incidence and reflection $y>y_2(x)$
is vacuous. The coating region $y_1(x) < y < y_2(x)$ is 
filled by a linear, homogeneous, isotropic, lossless dielectric--magnetic material characterized by relative permittivity $\epsilon_r$ and relative permeability  $\mu_r$. 
The grating surface $y=y_1(x)$ is periodic such that $y_1(x\pm d) = y_1(x)$
with $d$ as the period,
and the coating thickness $D$ is uniform so that $y_2(x) \equiv y_1(x)+D$. The grooves
of the grating are along the $z$ axis.
A linearly polarized electromagnetic plane wave of angular frequency $\omega$ is incident from the vacuous region at an angle $\theta_0$ with respect to the $y$ axis,
with the plane of incidence being the $xy$ plane. An $\exp(-i\omega t)$ 
dependence on time $t$ is implicit here.

Due to a perfectly conducting flat surface
coated with a uniform, lossless PPV material, the reflection is completely specular and the
specular reflectance possesses a magnitude of unity. If the signs of $\epsilon_r$ and $\mu_r$ are changed, the reflection is still specular, the magnitude of the specular reflectance still remains unity, but the phase of the reflection
coefficient changes \cite{Lem03}.

When the perfectly conducting surface is corrugated periodically, nonspecular modes
must arise in the reflected field. The characteristics of these modes will depend, in principle, on whether the coating material is of the PPV or of the NPV type. In
accordance with the focus of this communication, we must investigate two cases.
With the definition of the refractive index
\begin{equation}
n=\sqrt{\epsilon_r}\,\sqrt{\mu_r}\,,
\end{equation}
wherein the right side must not be confused with $\sqrt{\epsilon_r\mu_r}$,
these two cases are as follows:
\begin{itemize}
\item[I.] $n>0$, i. e., the coating material is of the PPV type;
\item[II.] $n<0$, i. e., the coating material is of the NPV type.
\end{itemize}

As mentioned in the previous section, we used a computer code
based on the differential or the C method, the details of which are available
elsewhere  \cite{chande,madeeasy,chandea2,DLopt}. This code yielded the reflection
coefficients $r_m$, $(m=0,\pm 1,\pm2\,...)$, wherefrom the reflection efficiencies
$e_m^r$ were computed \cite[Eq. 28]{DLopt} for the specular ($m=0$) and the nonspecular ($m\ne 0$) of the reflected field. Although our computational method
is suitable for grooves of arbitrary shape, we chose to restrict our attention here
to sinusoidal gratings; i.e.,
\begin{equation}
y_1(x)=0.5\,h\,\cos\left(\frac{2\pi x}{d}\right)\,, 
\end{equation}
where $h$ is the groove depth.

\section{Numerical results and discussion}
Calculations were made for the following constitutive parameters
of the coating material:
\begin{itemize}
\item[] Case I:  $\left\{\epsilon_r=2.5,\mu_r=1.2\right\}\Rightarrow n = 1.732$;
\item[] Case II: $\left\{\epsilon_r=-2.5,\mu_r=-1.2\right\}\Rightarrow n =- 1.732$.
\end{itemize}
Both linear 
polarization states~---~parallel $(p)$ and perpendicular $(s)$~---~of the
incident plane wave.

\begin{figure}[ht] 
\vspace{2cm}
\begin{center} 
\begin{tabular}{c}
\includegraphics[width=9cm]{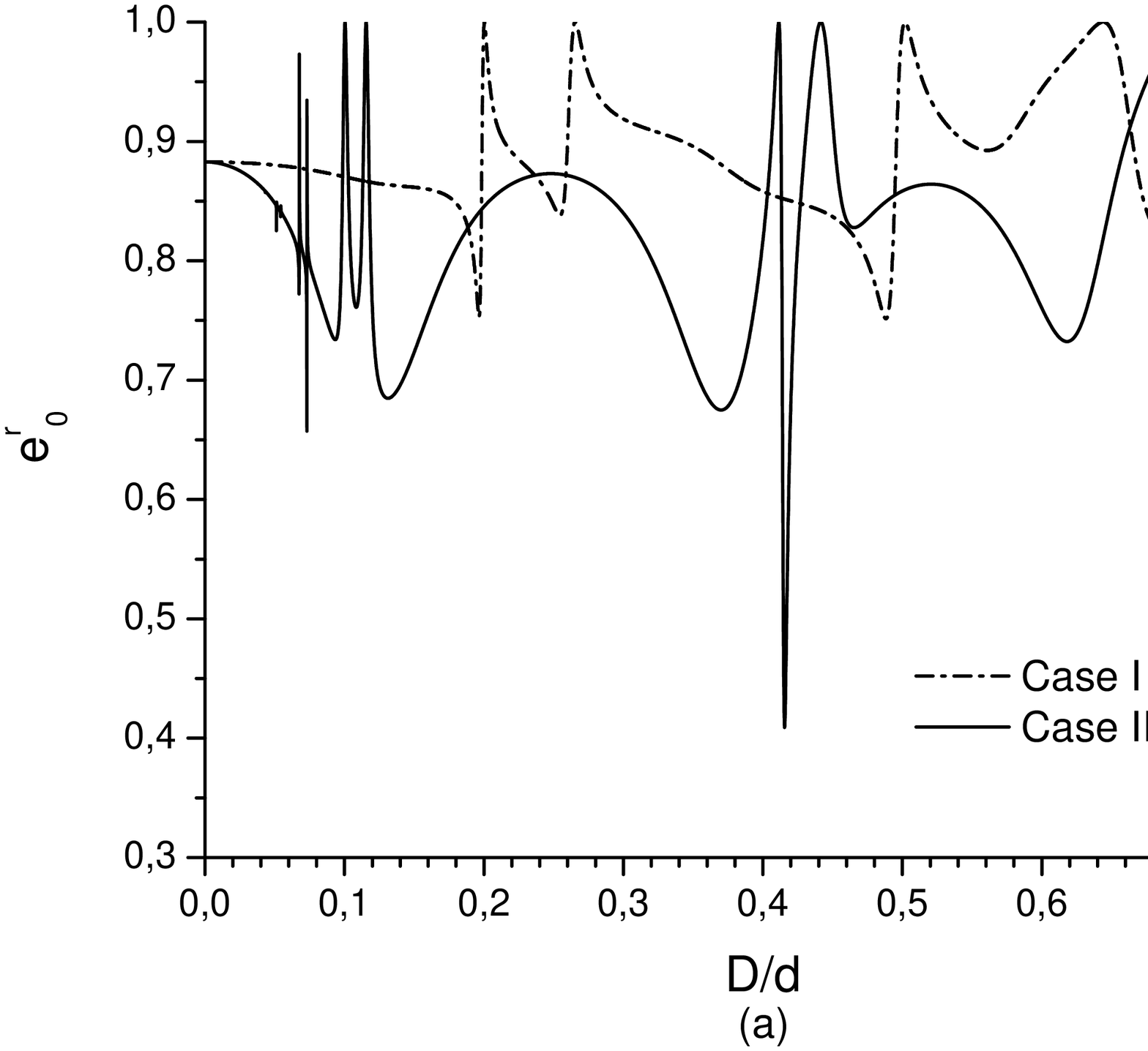} \hspace{-2cm} 
\includegraphics[width=9cm]{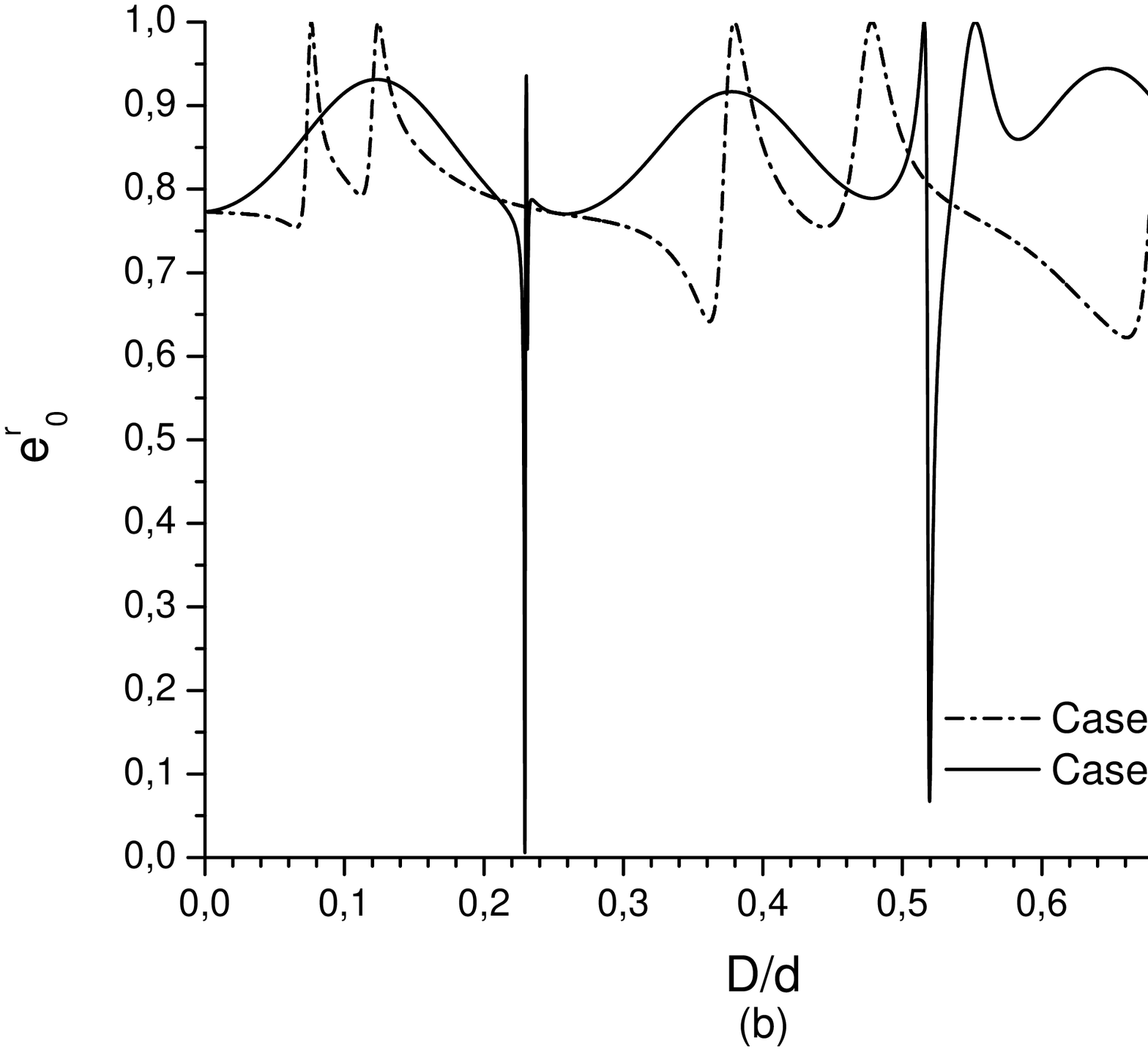} 
\end{tabular}
\end{center} 
\vspace{-2cm}
\caption[example]{Specular reflection efficiencies $e_0^r$ vs.
normalized thickness $D/d$ of the coating, when
$\theta_0=30^{\circ}$, $\lambda=0.81d$, and $h=0.1d$.
(a) $s$ polarization state, (b) $p$ polarization state.}
\label{Fig2}
\end{figure}

Let us begin with the specular reflection efficiencies $e_0^r$ calculated as functions of the
normalized coating thickness $D/d$, when the angle of incidence $\theta_0=30^\circ$,
the free--space wavelength $\lambda=0.81 d$, and the groove depth
$h=0.1d$. As may be gleaned from Fig.~\ref{Fig2},
 the specular reflection efficiencies are significantly affected by the type of the coating material. In particular, for the PPV coating material and for both $s$ and $p$ polarization states, the specular reflection efficiencies are high: more than 70\% of the energy incident onto the grating for the $s$ polarization state, and more than 60\% for the $p$ polarization state,
 is reflected specularly. Narrow dips appear at certain values of $D/d$, at which the value of the specular reflection efficiency decreases to a minimum value between 0.62 and 0.75, for both incident polarization states. Although  the specular
 reflection efficiencies are also high for the NPV coating material, the
 plots exhibit more pronounced dips. For example, the value of $e_0^r$  decreases abruptly for 
 the $s$ polarization state, reaching a minimum value of about 0.4 for $D/d \approx 0.42$; for the $p$ polarization state, two dips appear at $D/d \approx 0.22$ and $D/d \approx 0.52$, the specular reflection efficiency decreasing to about 0 and 0.1, respectively. 

\begin{figure}[hbt]
\vspace{2cm} 
\begin{center} 
\begin{tabular}{c}
\includegraphics[width=9cm]{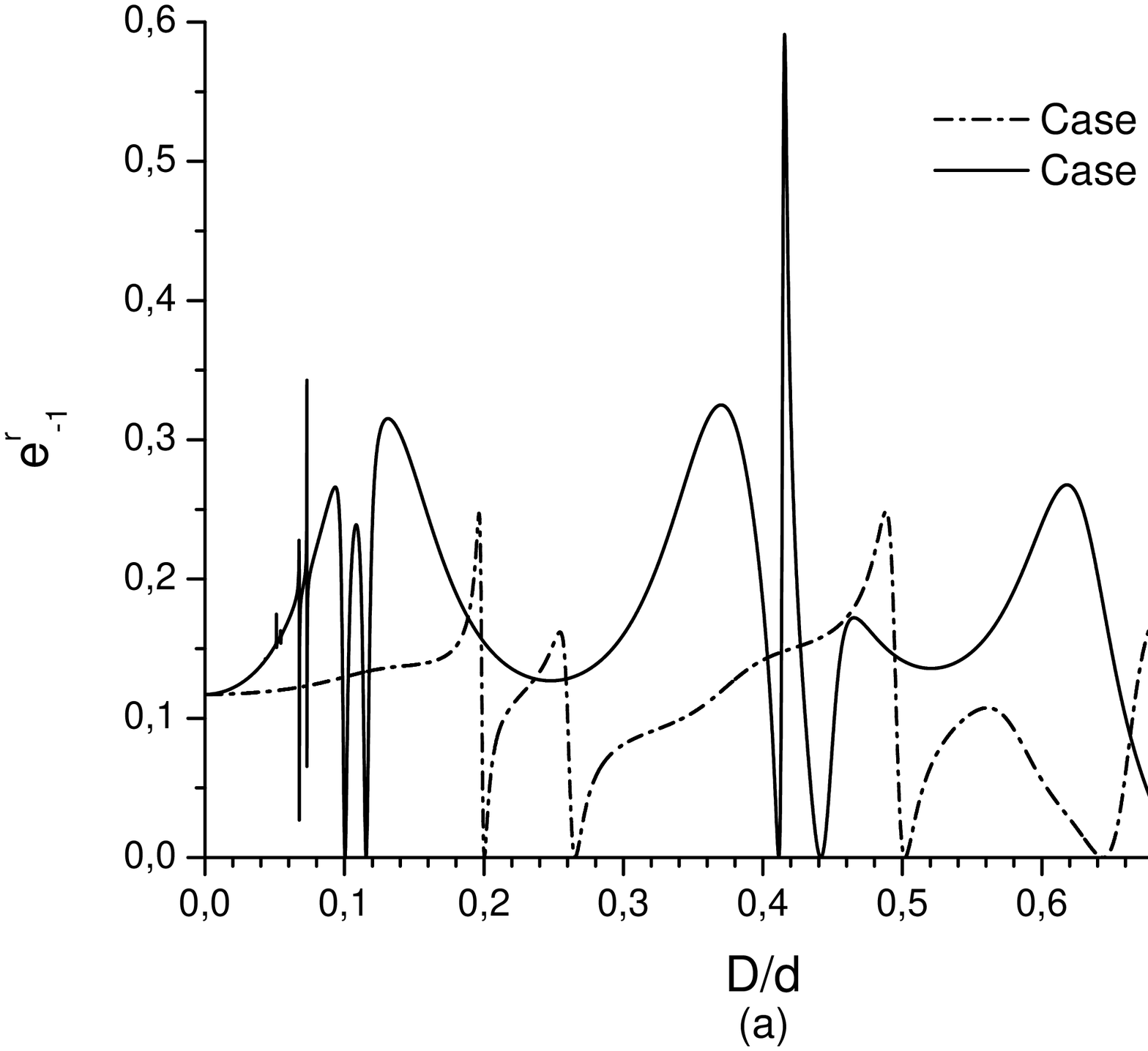} \hspace{-2cm} 
\includegraphics[width=9cm]{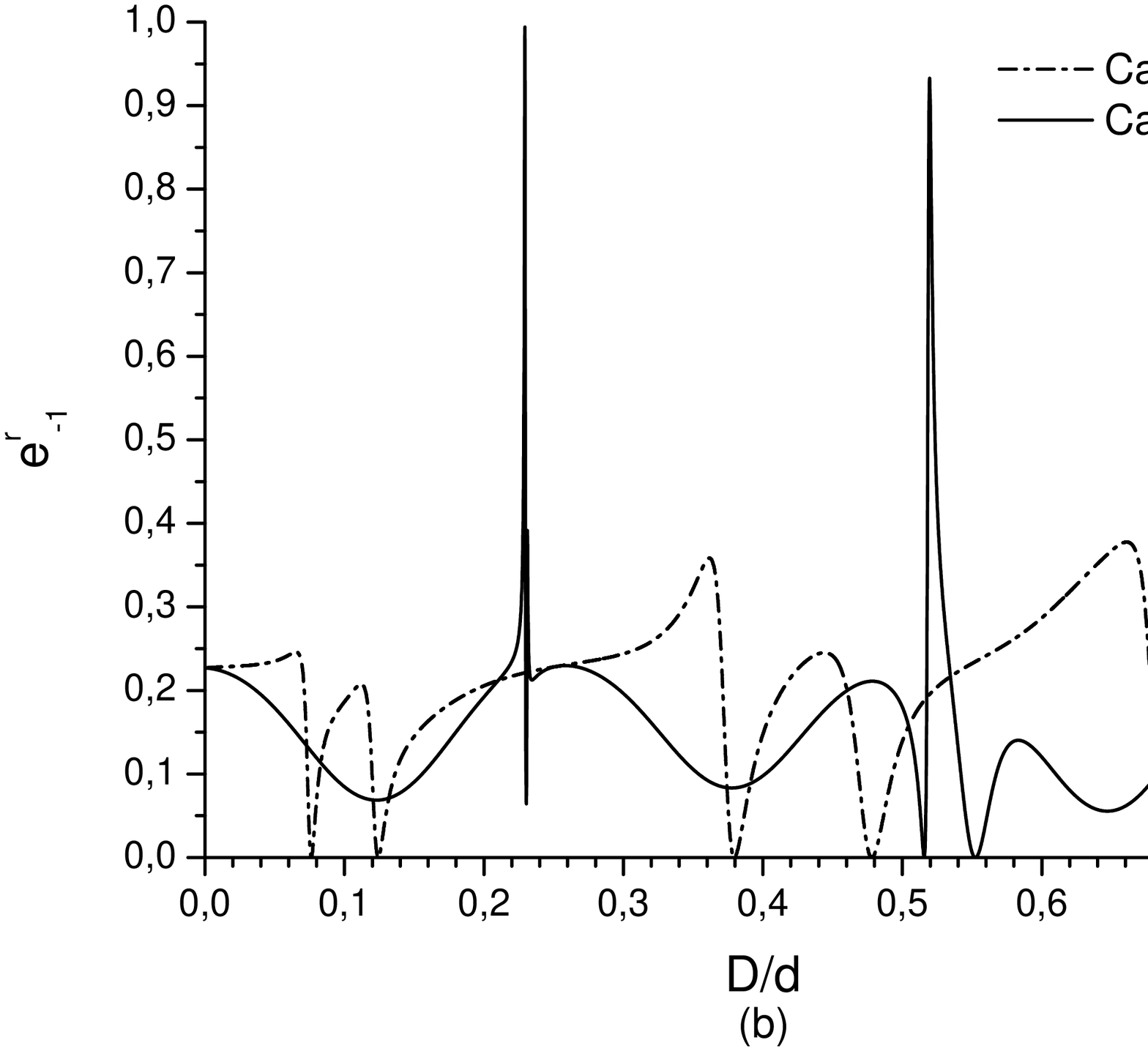}
\end{tabular}
\end{center} 
\vspace{-2cm}
\caption[example]{Same as Fig.~\ref{Fig2}, but for $e_{-1}^r$.}
\label{Fig3}
\end{figure}

When the value of $e_0^r$ dips, an enhancement of the energy carried by the $m=-1$ reflected order appears for both types of coating materials. However, the reflection of power into this order for the NPV coating material is much larger than for the
PPV coating material, a fact that can be appreciated in Fig.~\ref{Fig3}, wherein $e^r_{-1}$ is plotted against $D/d$ for both polarization states and the same parameters
as for Fig.~\ref{Fig2}. The increased maximum  nonspecular reflection efficiencies represent an important distinction between the PPV and NPV coating materials.

\begin{figure}[hbt] 
\vspace{2cm}
\begin{center} 
\begin{tabular}{c}
\includegraphics[width=9cm]{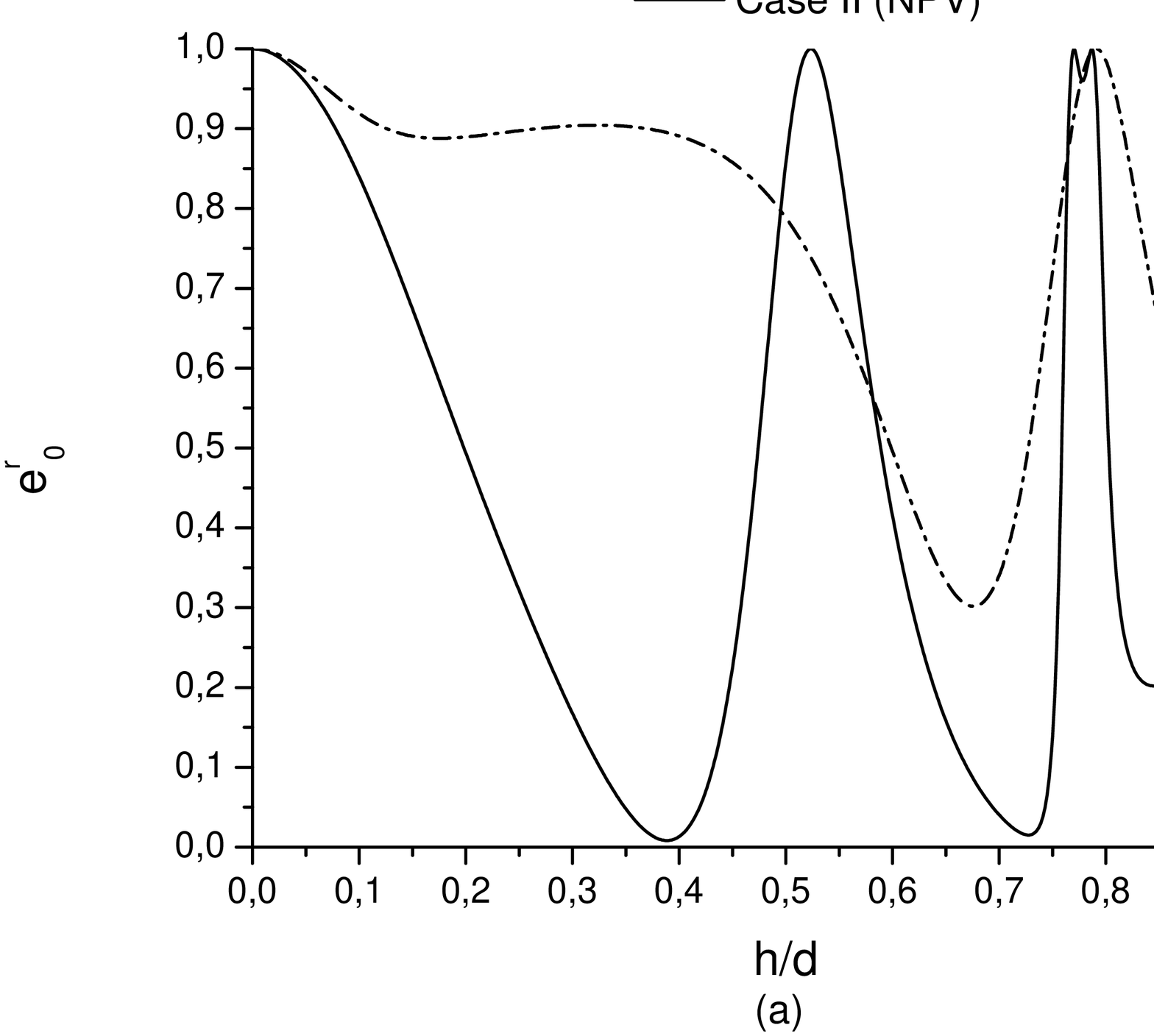} \hspace{-1cm} 
\includegraphics[width=9cm]{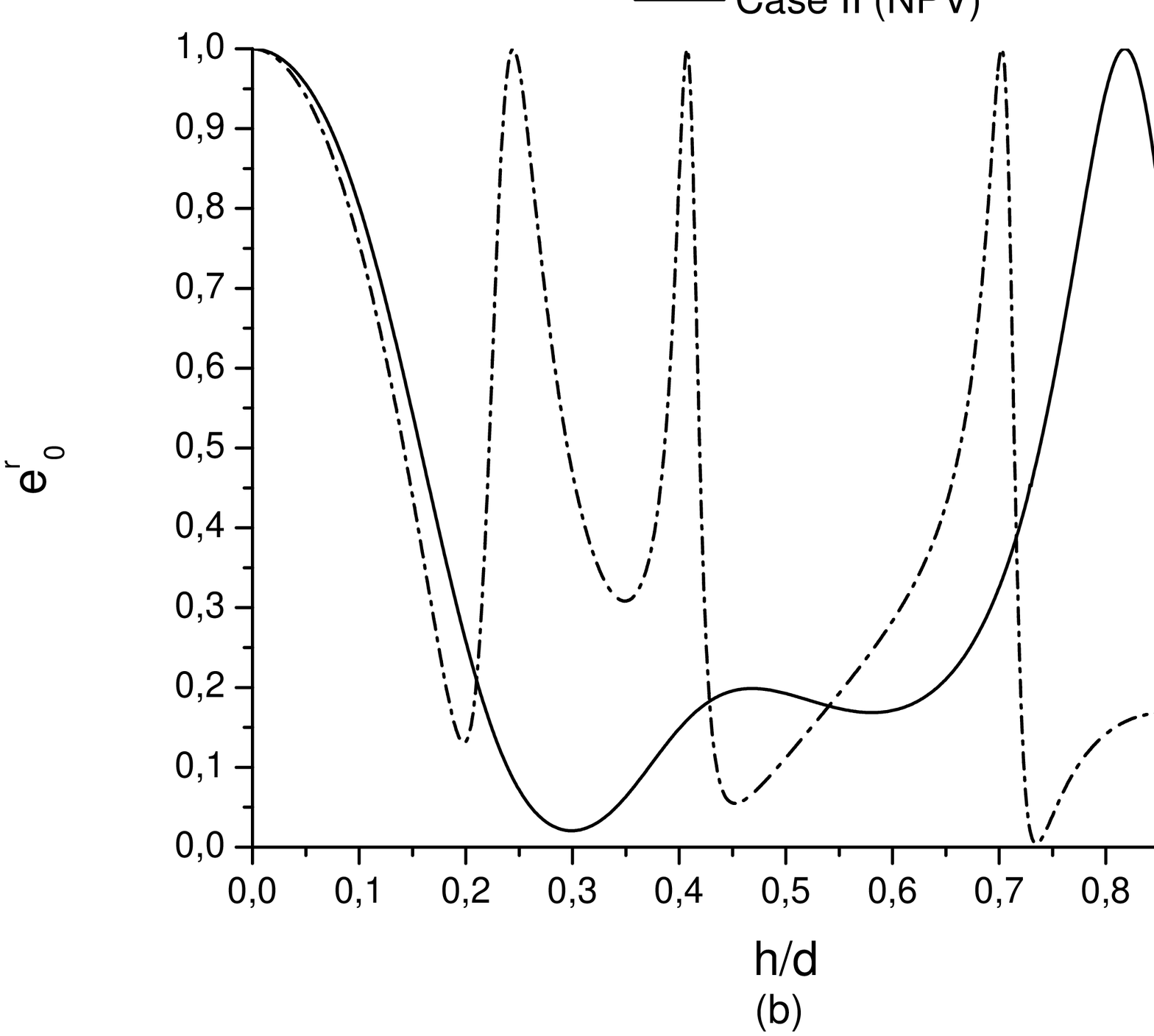}
\end{tabular}
\end{center} 
\vspace{-2cm}
\caption[example]{Specular reflection efficiencies $e_0^r$ vs.
normalized groove depth $h/d$ of the coating, when
$\theta_0=30^{\circ}$, $\lambda=0.81d$, and $D=0.3d$.
(a) $s$ polarization state,
(b) $p$ polarization state.}
\label{Fig4}
\end{figure}

Next we present the specular reflection efficiencies $e_0^r$ in Fig.~\ref{Fig4} as functions of the normalized groove depth
$h/d$ when the normalized coating thickness $D/d=0.3$, but the other parameters are the same as for Fig.~\ref{Fig2}. Let us begin analyzing the results for the $s$ polarization
state presented in Fig.~\ref{Fig4}(a). For the PPV coating material and for $h/d < 0.4$, $e^r_0$ does not vary significantly. However, for deeper grooves, the value of the 
specular reflection efficiency depends strongly on $h/d$: it first decreases to a minimum value of about $0.3$ for $h/d \approx 0.67$, then increases to unity,
and then decreases again. In contrast, for the NPV coating materials, significant fluctuactions in the value of $e_0^r$ are observed for the entire range of $h/d$:
$e_0^r\approx 0$ for certain values of $h/d$ ($\approx 0.38, 0.72$),  whereas 
$e_0^r\approx 1$ for other values of $h/d$ (e.g., $h/d \approx 0.52$).

\begin{figure}[hbt] 
\vspace{2cm}
\begin{center} 
\begin{tabular}{c}
\includegraphics[width=9cm]{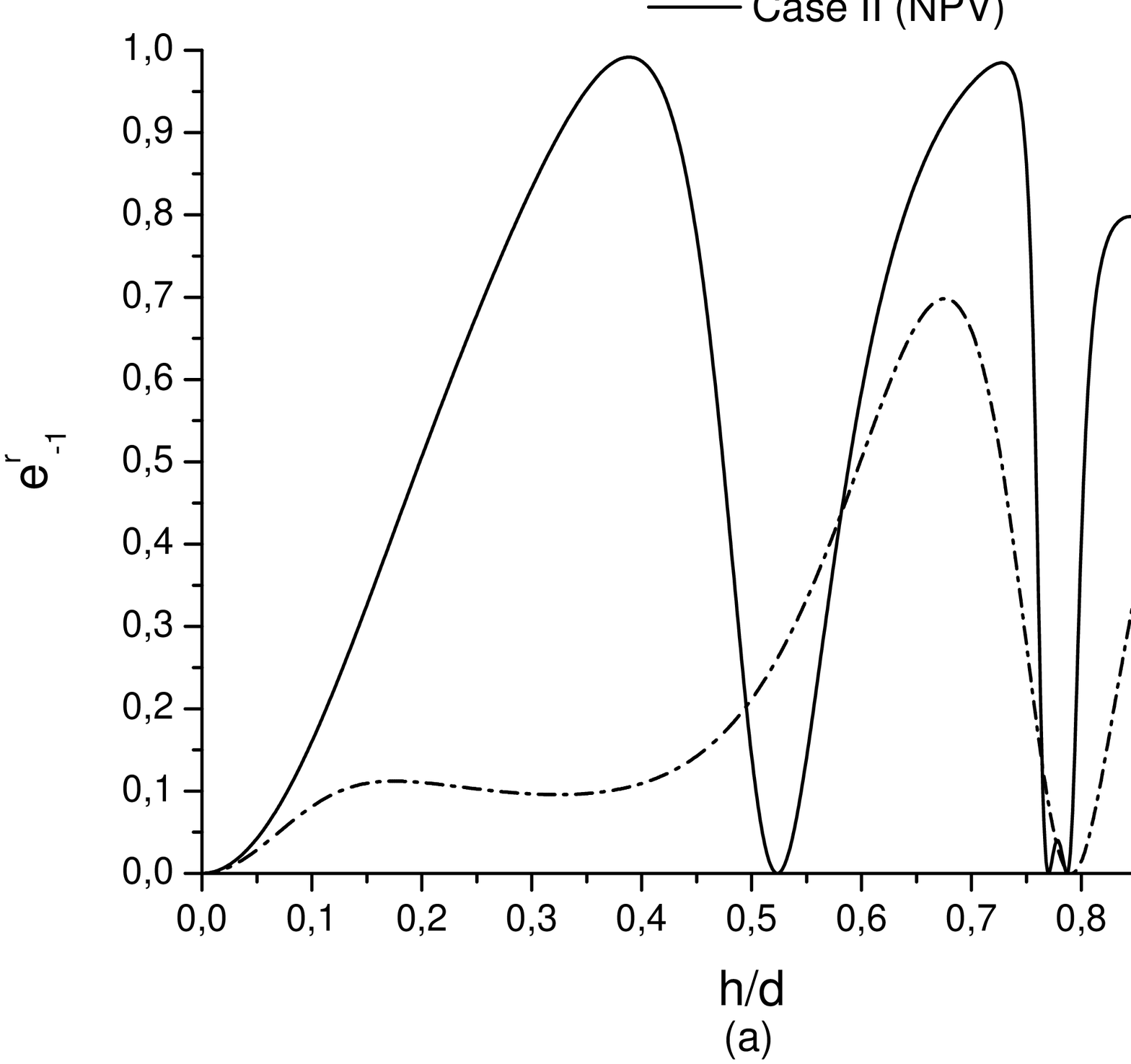} \hspace{-1cm} 
\includegraphics[width=9cm]{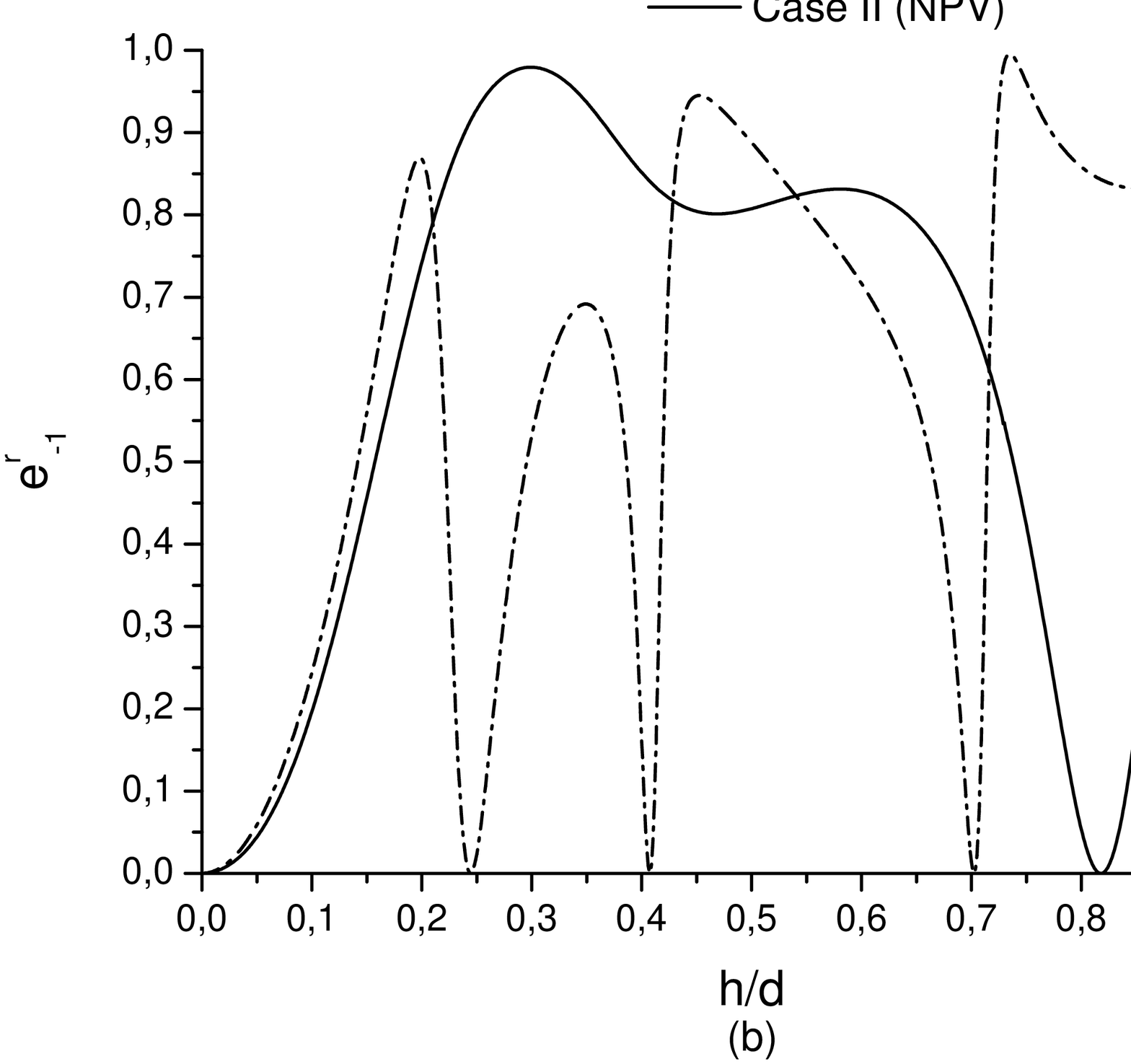}\
\end{tabular}
\end{center} 
\vspace{-2cm}
\caption[example]{Same as Fig.~\ref{Fig4}, but for $e_{-1}^r$.}
\label{Fig5}
\end{figure}

The $e_0^r$ vs. $h/d$ plots for the $p$ polarization state in Fig.~\ref{Fig4}(b)
are completely different from those for the $s$ polarization state  in Fig.~\ref{Fig4}(a).
The plot for the PPV coating material in Fig.~\ref{Fig4}(b) is qualitatively similar to the one for the NPV 
coating material in Fig.~\ref{Fig4}(a): large variations in the value of the specular reflection efficiency can be appreciated in the two plots. Conversely, for the NPV coating, the specular reflection efficiency
in Fig.~\ref{Fig4}(b) is generally quite low, except in the limit of the grating grooves becoming planar or when $h/d \approx 0.81$, where the entire incident energy is specularly reflected.
In Fig.~\ref{Fig5},   the nonspecular efficiencies of the $m=-1$ reflected order
are plotted as functions of $h/d$ for the same parameters as for Fig.~\ref{Fig4}.  An enhancement of the energy carried by the nonspecularly reflected order appears for both types of coating materials.

\begin{figure}[hbt] 
\vspace{2cm}
\begin{center} 
\begin{tabular}{c}
\includegraphics[width=9cm]{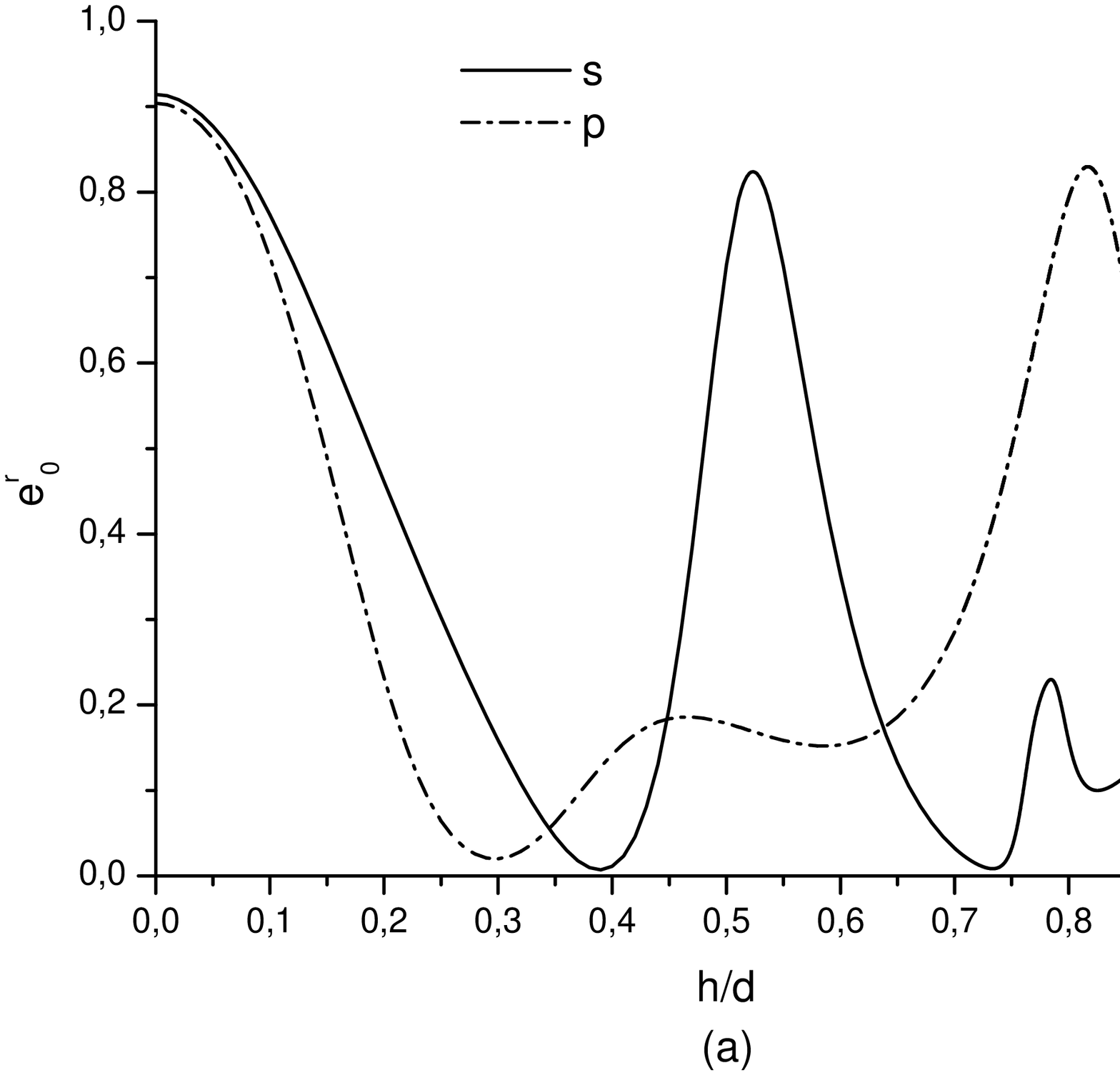} \hspace{-1cm} 
\includegraphics[width=9cm]{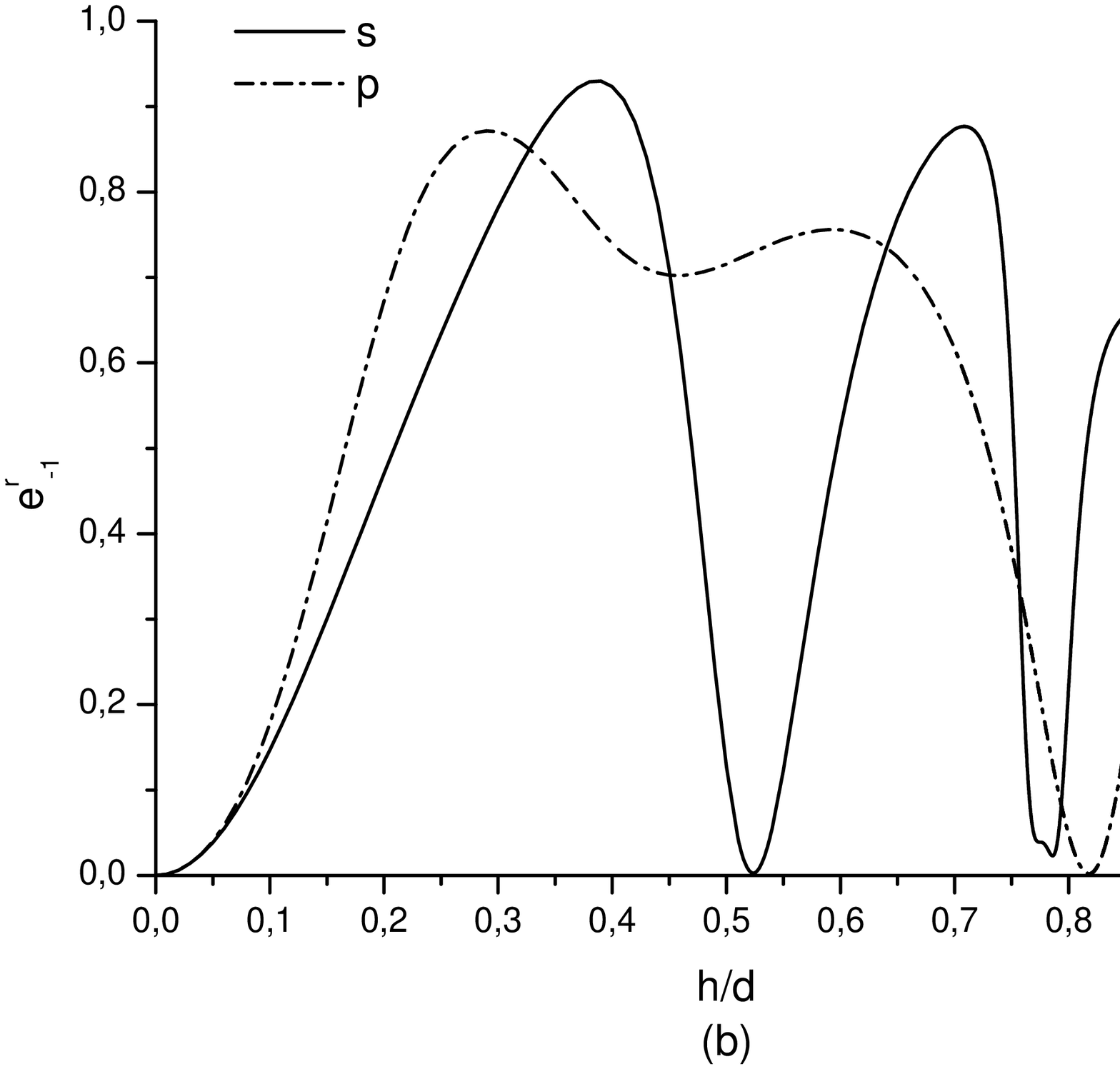}
\end{tabular}
\end{center} 
\vspace{-2cm}
\caption[example]{Reflection efficiencies  vs.
normalized groove depth $h/d$ of the coating, when
 $\epsilon_r = -2.5+i0.01$, $\mu_r=-1.2+i0.01$,
$\theta_0=30^{\circ}$, $\lambda=0.81d$, and $D=0.3d$.
(a) $e_0^r$,
(b) $e_{-1}^r$.}
\label{Fig6}
\end{figure}

All known NPV materials are both dispersive and absorptive. Dispersion
is irrelevant to this paper, as we have considered only a single frequency.
In order to identify the possible influence of weak absorption,
we set $\epsilon_r = -2.5+i0.01$ and $\mu_r=-1.2+i0.01$ for
the NPV coating material, and computed the  reflection
efficiencies $e_0^r$ and $e_{-1}^r$ for both polarization states 
as functions of the normalized
groove depth $h/d$, when $D/d=0.3$, $\lambda/d= 0.81$,
and $\theta_0=30^\circ$. Figure Fig.~\ref{Fig6} contains the resulting curves.
These are very similar to those for Case II in Figs.~\ref{Fig4} and 
\ref{Fig5} except
the efficiencies are somewhat lower. Thus, the introduction of weak
absorption in the NPV coating material is not likely to have a significant effect 
on the performance of the coated grating. 

\section{Conclusion}

Using a rigorous electromagnetic method, we investigated the response of conformally coated, perfectly conducting, sinusoidal gratings. We considered transparent coatings and   focused on the different responses obtained when the refractive index of the coating material is transformed from positive to negative. Our numerical results show that this transformation may have a significant influence on the diffraction efficiencies, a result that indicates that the emergence of homogeneous NPV materials promises new types of grating coatings which could be significantly different from their PPV counterparts. More specifically, coating materials with negative index of refraction 
may deliver enhanced maximum nonspecular reflection efficiencies in comparison
to coating materials with positive index of refraction.
Let us, however, caution that the foregoing conclusion requires confirmation by a
comprehensive investigation that we shall undertake next.

\newpage

\newpage

\newpage

\newpage

\newpage

\newpage

\end{document}